\documentclass[%
 reprint,
showpacs,
 amsmath,amssymb,
 aps,
pre,
]{revtex4-1}

\usepackage{graphicx}
\usepackage{dcolumn}
\usepackage{bm}

\newcommand{\fat}[1]{\mathbf{#1}}
\newcommand{\fatsym}[1]{\boldsymbol{#1}}
\newcommand{\D}{\mbox{d}}

\newcommand{\delt}{\Delta t}     
\newcommand{\te}[1]{\mathrm{#1}} 
\newcommand{\fe}{\fat{e}}

\begin{document}

\preprint{APS/123-QED}

\title{Rotational ratchets with dipolar interactions}

\author{Sebastian J\"ager}
 \email{jaeger@itp.tu-berlin.de}
\author{Sabine H. L. Klapp}%
\affiliation{%
Institute of Theoretical Physics, Technical University Berlin, \\
Hardenbergstr.~36, 10623 Berlin, Germany 
}%


\date{\today}

\begin{abstract}
We report results from a computer simulation study on the rotational ratchet
effect in systems of magnetic particles interacting via dipolar interactions.
The ratchet effect consists of directed rotations of the particles in an
oscillating magnetic field, which lacks a net rotating component. Our
investigations are based on Brownian dynamics simulations of such many-particle
systems. We investigate the influence of both, the random and deterministic
contributions to the equations of motion on the ratchet effect. As a main
result, we show that dipolar interactions can have an enhancing as well as a
dampening effect on the ratchet behavior depending on the dipolar coupling
strength of the system under consideration. The enhancement is shown to be
caused by an increase in the effective field on a particle generated by
neighboring magnetic particles, while the dampening is due to restricted
rotational motion in the effective field. Moreover, we find a non-trivial
influence of the short-range, repulsive interaction between the particles.
\end{abstract}

\maketitle


\section{Introduction}
\label{sec:introduction}
Thermal Brownian ratchets or, as they are sometimes called, Brownian motors,
are devices that are able to extract directional motion from Brownian random
noise \cite{haenggi2009}. In these out-of-equilibrium systems, it is possible
to rectify the Brownian noise into directional motion. In thermal equilibrium
such a phenomenon cannot exist as the second law of thermodynamics would be
violated \cite{reimann2002,astumian2002}.

Thermal ratchet effects have been known for a long time. Recently, however,
they are again gaining attention due to their possible applications in
biological transport \cite{magnasco1993, julicher1997} and nanotechnology
\cite{rousselet1994,linke1999}.

Most of the research on Brownian motors has been focused on directed {\it
translational} motion. Exceptions are recent studies of the so-called
rotational ratchet effect in ferrofluids, which has been investigated
theoretically \cite{engel2003,engel2004,becker2007} as well as experimentally
\cite{john2009}.

Ferrofluids are suspensions of ferromagnetic colloidal particles (with
diameters of about $10$nm or larger) in a carrier fluid such as water or oil
\cite{blum,rosenzweig}. These systems can be driven out of equilibrium by,
e.g., an oscillating magnetic field. The ratchet effect reported in
\cite{engel2004,engel2003} consists of a noise-driven directed rotation of the
particles, which are exposed to a field without a net rotating component. The
rotations of the particles are associated to an effective torque, which is
transferred to the solvent medium. This latter torque is of macroscopic size,
making the ferrofluid ratchet effect experimentally detectable \cite{john2009}.

The theoretical investigations so far have been performed on the basis of the
single-particle Langevin- and Fokker-Planck equations
\cite{engel2004,becker2007}. Interactions between the ferromagnetic colloids
have mostly been neglected \cite{engel2004}, the argument being that the
concentration of magnetic particles is extremely small in many ferrofluids
(volume fraction $\approx$ $1\%$). In more concentrated samples, however, one
would expect the magnetic dipole-dipole interactions between the particles to
become important. Indeed, a well known effect is the chain formation of the
particles triggered by the anisotropy, particularly the head-to-tail
preference, of the dipole interactions. There is, to our knowledge, only one
theoretical study in which the impact of the dipolar interactions on the
ratchet effect has been investigated \cite{becker2007}. This study approximates
the interactions on a mean-field level, i.e., all the particles experience a
homogeneous effective field.

In the present paper, we will investigate the impact of the {\it true} dipolar
interactions on a particle level, i.e., by Brownian dynamics (BD) computer
simulations. In this way we can not only capture the full anisotropy and range
of the interaction (which is known to be crucial for self-organization
processes in dipolar systems \cite{jaeger_klapp,jaeger2012,Jordanovic2011}),
but also the fact that the particles are mobile.

This paper is organized as follows: In Sec.~\ref{sec:model}, we present the
model and the simulation methods used throughout this study. The next section
deals with the rotational thermal ratchet effect in non-interacting systems.
Here, we will investigate the angular trajectories of the particles and the
influence of the strength of the noise and the external field. In
Sec.~\ref{sec:dipole} we will then turn to systems, in which the particles
interact via a short-range repulsive and a dipole-dipole potential. We will
show that dipolar interactions can enhance as well as suppress the ratchet
effect and we will analyze the mechanisms behind these effects. Further, we
show that the short-range isotropic, repulsive potential has a significant
influence on the ratchet behavior. The paper is then closed with a brief
summary and conclusions.

\section{Model and simulation methods}
\label{sec:model}
In this study, we consider a three-dimensional system of dipolar colloidal
particles that are immersed in a solvent. Only the dipolar particles are
handled explicitly. As a model we use a dipolar soft sphere (DSS) potential,
which is comprised of a repulsive part $U^{\mathrm{rep}}$ and a point
dipole-dipole interaction part $U^{\mathrm{D}}$:
\begin{equation}
    \label{eq:interaction}
    U^{\mathrm{DSS}}(\fat{r}_{ij}, \fatsym{\mu}_i, \fatsym{\mu}_j) =
    U^{\mathrm{rep}}(r_{ij}) + U^{\mathrm{D}}(\fat{r}_{ij}, \fatsym{\mu}_i, \fatsym{\mu}_j)
\end{equation}
In Eq.~(\ref{eq:interaction}), $\fat{r}_{ij}$ is the vector between the
positions of the particles $i$ and $j$, $r_{ij}$ its absolute value, and
$\boldsymbol{\mu}_i$ is the dipole moment of the $i$th particle. The dipolar
and repulsive interaction potentials are given by
\begin{equation}
    \label{eq:dipole}
    U^{\mathrm{D}}(\fat{r}_{ij}, \fatsym{\mu}_i, \fatsym{\mu}_j)
    = - \frac{3 (\fat{r}_{ij} \cdot \boldsymbol{\mu}_i) (\fat{r}_{ij} \cdot 
    \boldsymbol{\mu}_j)}{r_{ij}^5} + \frac{\boldsymbol{\mu}_i \cdot 
    \boldsymbol{\mu}_j}{r_{ij}^3}
\end{equation}
and
\begin{equation}
    \label{eq:rep}
    U^{\mathrm{rep}}(r) = U^{\mathrm{SS}}(r) - U^{\mathrm{SS}}(r_c) 
    + (r_c - r) \frac{\D U^{\mathrm{SS}}}{\D r}(r_c),
\end{equation}
respectively. Here, $U^{\mathrm{rep}}$ is the shifted soft sphere potential,
where
\begin{equation}
    U^{\mathrm{SS}}(r) = 4 \epsilon \left( \frac{\sigma}{r_{ij}} \right)^{12}
\end{equation}
is the unshifted soft sphere (SS) potential for particles of radius $\sigma$.

We investigate the system by making use of Brownian dynamics (BD) simulations.
These are based on the translational and rotational Langevin equations
\cite{mueller2002, jaeger_klapp}, which are integrated twice over a time
interval that is larger than the inertial relaxation time and small compared to
the time on which the configuration changes \cite{dhont1996, ermak1978,
banchio2003}. This procedure results in the equations \cite{ermak1978,
meriguet, dickinson}
\begin{equation}
    \label{eq:trans_eom_int}
    \fat{r}_i(t + \delt) = \fat{r}_i(t) + \frac{1}{k_B T} D_0^{\te{T}} \fat{F}_i \delt
                          + \sqrt{2 D_0^{\te{T}} \delt} \fatsym{\xi}_i^t
\end{equation} 
and
\begin{multline}
    \label{eq:rot_eom_int}
    \fat{e}_i(t + \delt) = \fat{e}_i(t) \\
    + \frac{1}{k_B T} D_0^{\te{R}} \fat{T}_i \delt \times \fat{e}_i(t)
    + \sqrt{2 D_0^{\te{R}} \delt} \fatsym{\xi}_i^r \times \fat{e}_i(t),
\end{multline}
which form the basis of our BD simulations. Equations (\ref{eq:trans_eom_int})
and (\ref{eq:rot_eom_int}) correspond to solving the Langevin equations in the
overdamped limit. In the equations above, $\fat{e}_i = \fatsym{\mu}_i/\mu$ is
the orientation of particle $i$. The conservative forces and torques are given by
\begin{align}
    \fat{F}_i & = - \fatsym{\nabla}_{\fat{r}_i} \sum_{j \neq i} U^\mathrm{DSS} (\fat{r}_{ij}, \fatsym{\mu}_i, \fatsym{\mu}_j) \\
    \fat{T}_i & = \fat{T}_i^\mathrm{DSS} + \fat{T}_i^\mathrm{ext},
\end{align} 
where
\begin{align}
    \fat{T}_i^\mathrm{DSS} & = - \fatsym{\mu}_i \times \sum_{j \neq i} \fatsym{\nabla}_{\fatsym{\mu}_i} U^\mathrm{DSS} (\fat{r}_{ij}, \fatsym{\mu}_i, \fatsym{\mu}_j) \\
    \label{eq:Text}
    \fat{T}_i^\mathrm{ext} & = \fatsym{\mu}_i \times \fat{B}^\mathrm{ext}
\end{align} 
with an external field $\fat{B}^\mathrm{ext}$. In Eqs.~(\ref{eq:trans_eom_int})
and (\ref{eq:rot_eom_int}), $D_0^{\te{T}}$ and $D_0^{\te{R}}$ are the
translational and rotational diffusion constants, which are given by
\begin{align}
    D_0^{\te{T}} & = \frac{k_B T}{3 \pi \eta \sigma} \\
    D_0^{\te{R}} & = \frac{k_B T}{\pi \eta \sigma^3} ,
\end{align}
where $\eta$ is the viscosity of the solvent. The quantities $\fatsym{\xi}_i^t$
and $\fatsym{\xi}_i^r$ are Gaussian random variables that satisfy
\begin{align}
    \langle \fatsym{\xi}^t_i \rangle  = 0, \quad
    \langle \fatsym{\xi}^r_j \rangle & = 0, \\
    \langle \fatsym{\xi}^t_i \fatsym{\xi}^t_j \rangle = \fatsym{\delta}_{ij}, \quad 
    \langle \fatsym{\xi}^r_i \fatsym{\xi}^r_j \rangle &
    = \fatsym{\delta}_{ij}, \quad \langle \fatsym{\xi}^r_i \fatsym{\xi}^t_j \rangle  = 0.
\end{align} 

Regarding the external field $\fat{B}^\mathrm{ext}$, we use the same ansatz
suggested previously in Refs.~\cite{engel2004,becker2007}. Specifically, the
field has a constant component in $x$-direction and an oscillating, yet
asymmetric component in $y$-direction. A suitable ansatz is given by
\begin{equation}
    \label{eq:osc_field}
    \fat{B}^\mathrm{ext}(t) = B_x \fat{e}_x
    + B_y [\cos (\omega_0 t) + \sin (2 \omega_0 t + \delta)] \fat{e}_y  .
\end{equation}
The important point is that this field involves only oscillations, but no full
rotations, irrespective of the phase shift $\delta$. Nevertheless, it turns out
that the particles can perform directed full rotations. We note that the ansatz
(\ref{eq:osc_field}) is, however, by no means the only field with which a
ratchet effect can be realized. In fact a multitude of different fields are
suitable, if certain certain conditions are met: $B_x$ must be non-vanishing
and there cannot be a $\Delta t$ such that $B_y(t) = B_y(-t + \Delta t)$ (see
Ref.~\cite{engel2004} for a detailed discussion of this issue).

Note that the particles we consider here are immersed in a solvent. However,
the solvent is only taken into account implicitly and on a single particle
level, i.e., the random noise and the diffusion constants do not depend on the
configuration of the particles.

We consider $N = 500$ particles in our simulation box with periodic boundary
conditions. The long-range dipolar interactions are taken into account by using
the Ewald summation method \cite{Weis2003}.

For convenience, we make use of the following reduced units: Field strength
$B_\alpha^* = (\sigma^3/\epsilon)^{1/2} B_\alpha$ ($\alpha = x, y$); dipole
moment $\mu^* = (\epsilon \sigma^3)^{-1/2} \mu$; torque $\fat{T}^* =
\fat{T}/\epsilon$; time $t^* = t D_0^\mathrm{T}/\sigma^2$; temperature $T^* =
k_B T/\epsilon$; position $\fat{r}^* = \fat{r}/\sigma$.

In addition, we will employ the parameter $\lambda = \mu^{*2}/T^*$ measuring
the dipolar coupling strength relative to $k_B T$.

\section{Results}
\subsection{The thermal ratchet effect in a non-interacting system}
\label{sec:ratchet}
As a background for our investigation of the impact of dipolar interactions, we
discuss in this section BD simulation results for the rotational thermal
ratchet effect in systems of non-interacting particles. To this end, we analyze
the trajectories of the particles under the influence of the external field
(\ref{eq:osc_field}) as well as the corresponding torque. Further, we
investigate the dependence of the ratchet effect on the strength of the
external field versus that of the noise. We note that some of the points
discussed in this section have already been investigated in
Ref.~\cite{engel2004} via numerical integration of the respective Fokker-Planck
equation. Our present BD results supplement these previous theoretical results.

The systems we consider in this section are characterized by a temperature $T^*
= 0.2$. The particles are driven by a field of frequency $\omega
\sigma^2/D_0^\te{T} = 15$, $y$-field component $B_y^* = 1$, and various values
of $B_x^*$. The dipole moment is set to $\mu^* = 1$, such that the dipole-field
coupling is $\mu B_y/k_B T = \mu^* B_y^*/T^* = 5$.

Since neither repulsive nor dipolar interactions are taken into account in this
section, the density can be chosen arbitrarily. With these choices of the
variables, our results are easily comparable to the ones from
Ref.~\cite{engel2004} (for the precise relations between the dimensionless
variables in our study and those in \cite{engel2004}, see the Appendix).

To start with, we show in Fig.~\ref{fig:exey00} the mean orientation of the
particles,
\begin{equation}
    \label{eq:mean_orient}
    \bar{\fat{S}} (t) = \frac{1}{N} \sum_{i=1}^{N} \fat{e}_i (t),
\end{equation}    
for the external field (\ref{eq:osc_field}) with $B_x^* = 0.1$, $0.3$ and
$\delta = 0$. As can be seen, $\bar{S}_x$ is essentially constant, while
$\bar{S}_y$ follows (with a phase lag) the oscillating component of the
external field indicated by the dotted line in Fig.~\ref{fig:exey00}.
Interestingly, while $\bar{S}_x$ is increased for the field with $B_x^* = 0.3$
over the field with $B_x^* = 0.1$, $\bar{S}_y$ remains essentially unchanged.
We will later see that this is of crucial importance for understanding the
impact of interactions.
\begin{figure}[ht]
    \centering
    \includegraphics[width=83mm]{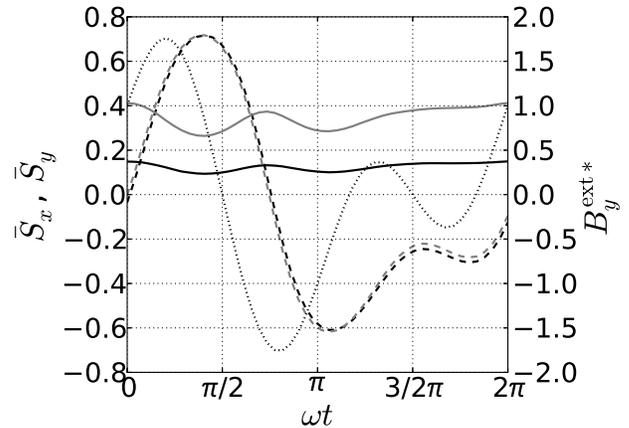}
    \caption{Mean orientation of the particles during one rotational period of
    the field for non-interacting systems with $B_x^* = 0.1$ (black lines) and
    $B_x^* = 0.3$ (gray lines). The $x$- and $y$-components of $\bar{\fat{S}}
    (t)$ are indicated by solid and dashed lines, respectively. The dotted line
    shows the field component in $y$-direction $B_y^{\mathrm{ext}*}$.}
    \label{fig:exey00}
\end{figure}

\begin{figure}[ht]
    \centering
    \includegraphics[width=83mm]{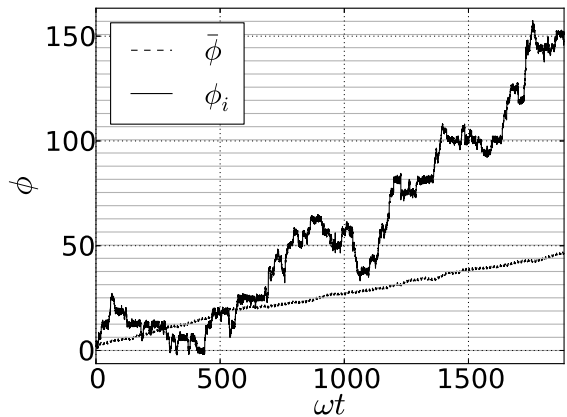}
    \caption{The angular trajectories in terms of the polar angle $\phi_i$ of
    an individual particle and the system-averaged angle $\bar{\phi}$ at $B_x^*
    = 0.3$ and $\delta = 0$.}
    \label{fig:time_phi}
\end{figure}

The behavior of the mean orientation seen in Fig.~\ref{fig:exey00} appears
essentially deterministic. The actual ratchet effect is illustrated in
Fig.~\ref{fig:time_phi}, where we plot two angles $\phi_i$ and $\bar{\phi}$.
The former is the angle that an (arbitrary) particle $i$ encloses with the
$x$-axis. It first remains close to a multiple of $2 \pi$ (indicated by the
horizontal lines) corresponding to the particle oscillating around the
$x$-direction of the field. This behavior continues, until a noise-induced full
rotation (i.e., a crossing of a horizontal line) occurs. One also sees that the
forward rotation, i.e., an increase by $2 \pi$, occurs more often than the
corresponding backward rotation. This illustrates the directional character of
the ratchet effect.

The second quantity $\bar{\phi}$ depicted in Fig.~\ref{fig:time_phi}
corresponds to the averaged value of the angles $\phi_i$ of all the particles
\begin{equation}
    \bar{\phi}(t) = \frac{1}{N} \sum_{j=1}^{N} \phi_j(t).
\end{equation}
In contrast to $\phi_i$ this average angle $\bar{\phi}$ increases
monotonically, since individual fluctuations are smeared out.

\begin{figure}[ht]
    \centering
    \includegraphics[width=83mm]{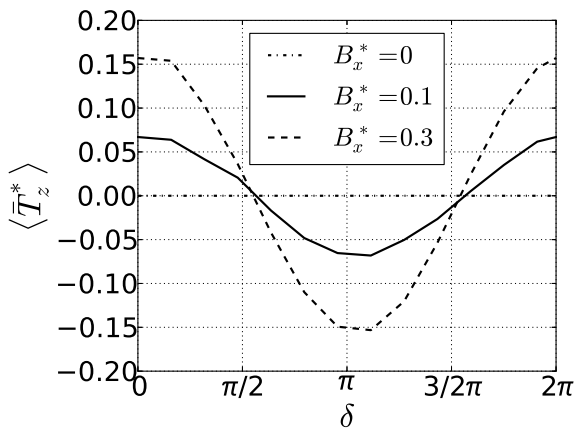}
    \caption{Averaged torque during one rotational period of the field as
    function of the phase difference $\delta$ for $B_x^* = 0, 0.1$, and $0.3$
    (and $B_y^* = 1$).}
    \label{fig:torques_delta}
\end{figure}

Irrespective of these differences between $\phi_i$ and $\bar{\phi}$,
Fig.~\ref{fig:time_phi} clearly demonstrates that there is a net rotational
motion in one direction. This corresponds to the presence of a net torque. We
calculated the net torque as an average of the time-dependent torque over one
period of the field, i.e.,
\begin{equation}
    \langle \bar{\fat{T}} \rangle = \frac{1}{\tau} \int_{t_0}^{t_0 + \tau} \bar{\fat{T}} (t) \D t  
\end{equation}   
where
\begin{equation}
    \label{eq:ttime}
    \bar{\fat{T}}(t) = \frac{1}{N} \sum_{i=1}^{N} \fatsym{\mu}_i(t) \times \fat{B}_i(t).
\end{equation} 
with
\begin{equation}
    \fat{B}_i = - \nabla_{\fatsym{\mu}_i} \sum_{j \neq i}
    U^\mathrm{DSS} (\fat{r}_{ij}, \fatsym{\mu}_i, \fatsym{\mu}_j) + \fat{B}^\mathrm{ext} .
\end{equation}
Numerical data for the net torque that the particles experience for $B_x^* =
0$, $0.1$, and $0.3$ over the phase difference $\delta$ is presented in
Fig.~\ref{fig:torques_delta} [cf.~Eq.~(\ref{eq:osc_field})].
It is seen that $\langle \bar{T}_z^* \rangle$ is indeed non-zero and (at
$\delta = 0$) positive, reflecting the net rotation of the particles to the
right. Increasing the phase difference $\delta$, the value of $\langle
\bar{T}_z^* \rangle$ changes and even assumes negative values. This implies
that the particles can perform forward as well as backward rotations depending
on $\delta$.

We now consider in more detail the dependence of the net torque on the strength
of the constant field contribution $B_x^*$. At $B_x^* = 0$, the external field
simply performs an oscillation into the $y$-direction. In that case, no net
torque can be observed. With only one direction distinguished by the field,
directional rotational motion simply cannot occur \cite{reimann2002}. For
non-vanishing $x$-components the magnitude of $\langle \bar{T}_z^* \rangle$
depends strongly on the value of $B_x^*$. As illustrated by
Fig.~\ref{fig:torques_delta}, increasing $B_x^*$ from $0.1$ to $0.3$ results in
considerably larger torques $\langle \bar{T}_z^* \rangle$. We explain this
increase as follows: At higher values of $B_x^*$, the particles are much more
aligned into the $x$-direction of the field, and thus in the plane of the
field. The latter point is illustrated by Fig.~\ref{fig:exey00}: $\bar{S}_x$ is
considerably larger for $B_x^* = 0.3$ than for $B_x^* = 0.1$ while $\bar{S}_y$
remains essentially unchanged. In other words, higher values of $B_x^*$ ensure
that the particles remain within the plane of the field without dampening the
oscillations of the dipole moments in the $y$-direction.

\begin{figure}[ht]
    \centering
    \includegraphics[width=83mm]{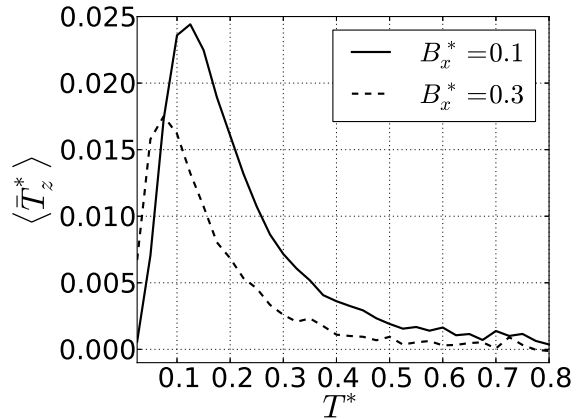}
    \caption{The net torque as function of the temperature for $B_x^* = 0.1$, $0.3$.}
    \label{fig:temp_torque}
\end{figure}

It is well established that ratchet effects, in general, depend strongly on the
strength of the noise relative to the deterministic contributions to the
equations of motion \cite{reimann2002}. For the present system, this interplay
is illustrated in Fig.~\ref{fig:temp_torque}, where we plot the torque as a
function of the dimensionless temperature $T^*$. Inspecting
Eq.~(\ref{eq:rot_eom_int}), we see that the temperature $T^*$ influences the
strength of the deterministic torque (due to the field) alone, if the diffusion
constant is kept constant. In other words, $T^*$ is a measure for the
aforementioned ratio of conservative torques to random noise. Small
temperatures correspond to systems that are dominated by deterministic torques,
while large temperatures correspond to noise-dominated systems.

In Fig.~\ref{fig:temp_torque}, we can see that the ratchet effect is strongest
for finite temperatures in the range $T^* \approx 0.05 - 0.2$. This means that
the ratchet effect looses in strength for too small or too large noise
contributions. If the temperature is too small, the field dominates the
rotational motion of the particles, which are effectively unable to perform
rotations against the field. At large temperatures, on the other hand, the
noise dominates such that the influence of the field becomes insignificant.
However, without the {\it non-equilibrium} influence of the external driving
field, the ratchet effect cannot exist \cite{reimann2002, astumian2002}.

In Ref.~\cite{engel2004}, the behavior of a single dipole in the oscillating
field (\ref{eq:osc_field}) was investigated on the basis of a Fokker-Planck
equation. Consistent with our results, the authors of Ref.~\cite{engel2004}
found a maximum in the strength of the net torque at finite values of the noise
intensity. Moreover, for the particular choice $B_x^* = 0.3$, our results for
the mean orientation and the net torque [see Figs.~\ref{fig:exey00} and
\ref{fig:torques_delta}] are in quantitative agreement with those in
Ref.~\cite{engel2004} (see the Appendix for the relation between the
dimensionless units).

\subsection{Influence of the particle interactions}
\label{sec:dipole}
In a ferrofluid, the particles interact with each other via short-range
repulsive as well as dipolar interactions. These interactions can be neglected
in strongly diluted ferrofluids, they do, however, become important when the
density of the dipolar particles becomes higher.

As has been shown in previous studies, particle interactions can indeed have a
profound influence on ratchet effects
\cite{reimann2002,pototsky2011,becker2007}: For instance, in a translational
ratchet, they can reverse the direction of the effect or even give rise to it
in the first place \cite{reimann2002,reimann1999}. The latter is also true for
the rotational ratchet effect. It was shown in Ref.~\cite{becker2007} that
dipolar interactions treated on a simple mean-field level can induce effective
particle rotations despite a vanishing field component $B_x$.

In the following, we choose a density of $\rho \sigma^3 = 0.2$ corresponding to
a dipolar fluid of moderate packing fraction $\eta = \pi \rho \sigma^3/6
\approx 0.105$ \cite{ivanov2001}. This choice ensures that the dipolar
interactions play a crucial role at higher coupling strengths. The frequency of
the oscillating field is again set to $\omega \sigma^2/D_0^\te{T} = 15$.
Regarding the interaction parameters, we consider a range of values for the
dimensionless dipole moment $\mu^*$ and different values of the dimensionless
temperature $T^*$. In this way we can explore both, impact of the dipolar
interactions (\ref{eq:dipole}), and those of the repulsive interactions
(\ref{eq:rep}). 

Note that while we vary $\mu^*$ (and, thus, $\lambda$), we keep the products
$\mu^* B^*_x$ and $\mu^* B^*_y$, i.e., the dipole-field coupling, fixed. To
indicate the used field strength, we therefore use the notation $B^+_\gamma
\equiv B^*_\gamma / \mu^*$.

In the lower temperature systems ($T^* = 0.2$) considered here, we use $B_y^+ =
1$. With this choice the interaction strength between dipoles and field remains
as in the previous section. In the systems with $T^* = 1$, we use
proportionally stronger external fields with $B_y^+ = 5$. The relative strength
of the external field compared to the Brownian noise is then equal to the one
in the low temperature case.

\begin{figure}[ht]
    \centering
    \includegraphics[width=83mm]{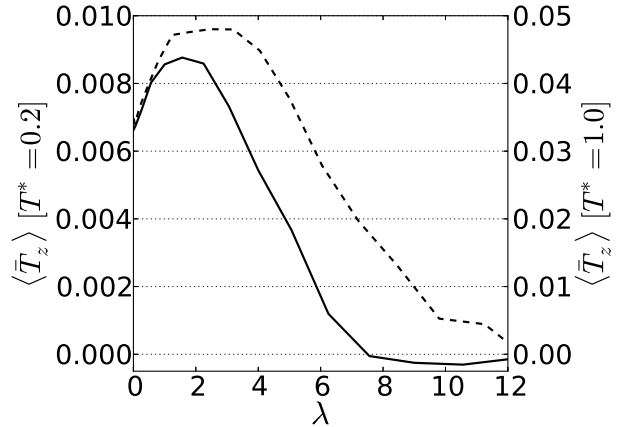}
    \caption{The z-component $\langle \bar{T}_z^* \rangle$ of the averaged
    torque for different coupling strengths $\lambda$ at $T^* = 1$, $B_x^+ =
    0.5$ (solid line, right axis), and $T^* = 0.2$, $B_x^+ = 0.1$ (dashed line,
    left axis).}
    \label{fig:phi_mu}
\end{figure}

In Fig.~\ref{fig:phi_mu} we present results for the z-component $\langle
\bar{T}_z^* \rangle$ of the averaged torque $\langle \bar{\fat{T}}^* \rangle$
for systems with different dipolar coupling strengths $\lambda = \mu^{*2}/T^*$
at $B_x^+ = 0.5$, $T^* = 1$ and $B_x^+ = 0.1$, $T^* = 0.2$. At $\lambda = 0$,
the particles interact with each other via the soft-sphere interaction but not
via the dipole-dipole interaction. Note that pure soft-sphere interactions do
not affect particle rotations, and thus they should not influence the ratchet
effect. Therefore the net torque found at $\lambda = 0$ for the $T^* = 0.2$
system equals the one shown in Fig.~\ref{fig:torques_delta} for $\delta = 0$.

Starting from the non-interacting system, the net torque plotted in
Fig.~\ref{fig:phi_mu} increases up to $\lambda \approx 1.5$ ($T^* = 1$) or
$\lambda \approx 2.5$ ($T^* = 0.2$), respectively. In the high temperature
system, the maximum of $\langle \bar{T}_z^* \rangle$ is approximately $30\%$
larger than the torque at $\lambda = 0$. For $T^* = 0.2$ the maximum is even
more pronounced: The torque is increased by nearly $40\%$ compared to the
non-interacting system. For higher values of $\lambda$, $\langle \bar{T}_z^*
\rangle$ decreases continuously.
\begin{figure}[ht]
    \centering
    \includegraphics[width=83mm]{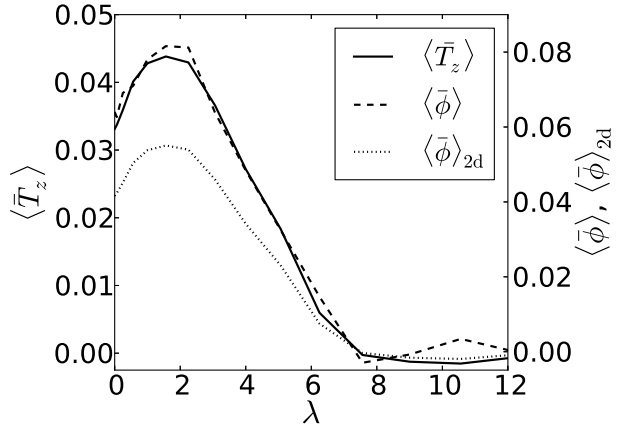}
    \caption{
    The net torque $\langle \bar{T}_z^* \rangle$, the mean traversed angle
    $\langle \bar{\phi} \rangle$, and the mean traversed angle $\langle
    \bar{\phi} \rangle_\mathrm{2d}$ calculated from the torque [see
    Eq.~(\ref{eq:phi_torque})] are shown for $T^* = 1$ and $B_x^+ = 0.5$.}
    \label{fig:phi_mu2}
\end{figure}

The behavior of $\langle \bar{T}_z^* \rangle$ is also reflected by the change
in the (system-averaged) polar angle
$\phi$
\begin{equation}
    \langle \bar{\phi} \rangle = \bar{\phi} (t + \tau) - \bar{\phi} (t)
\end{equation}
during one rotational period of the field. As shown in Fig.~\ref{fig:phi_mu2},
$\langle \bar{\phi} \rangle$ and $\langle \bar{T}_z^* \rangle$ behave almost
identically, which can be understood by looking at the BD evolution equation
(\ref{eq:rot_eom_int}). This equation corresponds to numerically integrating 
\begin{equation}
    \label{eq:rot_eom}
    \fatsym{\omega}_i = \frac{1}{k_B T} D_0^{\te{R}} \fat{T}_i + \sqrt{2 D_0^{\te{R}}} \fatsym{\zeta}_i.
\end{equation}
Here, $\fatsym{\omega}_i = \fe_i \times \dot{\fe}_i$ is the angular velocity of
particle $i$ and $\fatsym{\zeta}_i$ is a random Gaussian variable. Note that a
differential equation for $\dot{\fat{e}}_i$ can be obtained from
Eq.~(\ref{eq:rot_eom}) by using the definition of $\fatsym{\omega}_i$ and
taking the vector product with $\fat{e}_i$. Equation (\ref{eq:rot_eom_int})
then corresponds to integrating the resulting equation. Assuming the rotational
motion of the particles to be restricted to the plane of the field (denoted by
the subscript ``2d'') and neglecting the random noise, we find
\begin{equation}
    \label{eq:phi_torque}
    \langle \bar{\phi} \rangle_\mathrm{2d} = \frac{\tau D_0^\te{R}}{k_B T}
    \langle \bar{T_z} \rangle_\mathrm{2d}
\end{equation}
from integrating both sides of Eq.~(\ref{eq:rot_eom}) over one period of the
field. This equation relates the traversed angle of the particles to the
average torque. The dotted line in Fig.~\ref{fig:phi_mu2} demonstrates this
relation. We calculated the traversed angle via Eq.~(\ref{eq:phi_torque}) from
the torque component $\langle \bar{T}_z \rangle$. It is seen that this equation
slightly underestimates the observed value of $\langle \bar{\phi} \rangle$,
which can be explained by the fact that Eq.~(\ref{eq:phi_torque}) holds
strictly only for particle rotations in the plane of the field. Additionally,
we neglected the random noise, which is expected to introduce further
deviations from the observed relation. Nonetheless, Eq.~(\ref{eq:phi_torque})
captures the general behavior of $\langle \bar{T_z} \rangle$ as function of
$\lambda$ quite well.

\begin{figure}[ht]
    \centering
    \includegraphics[width=83mm]{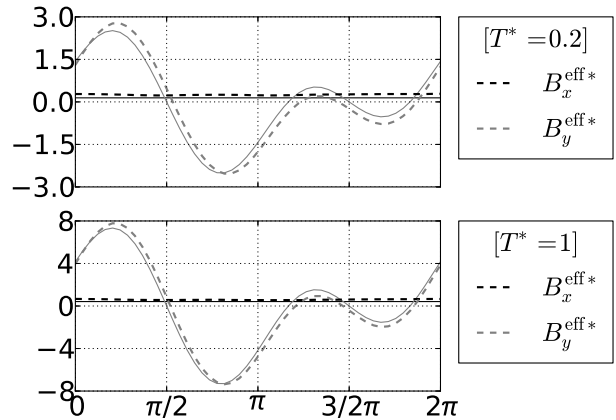}
    \caption{Mean values of the $x$- and $y$-components of the averaged
    effective field $\fat{B}^{\mathrm{eff}*}$ over one rotational period of the
    field at $\lambda = 2.45$, $T^* = 0.2$, and $B_x^+ = 0.1$ (top) and
    $\lambda = 1.44$, $T^* = 1$, and $B_x^+ = 0.5$ (bottom). The solid lines
    represent the respective external fields, i.e., $B^+_x = 0.1$, $B^+_y = 1$
    (top); $B^+_x = 0.5$, $B^+_y = 5$ (bottom)}.
    \label{fig:beff}
\end{figure}

We now discuss the origin of the maximum in Figs.~\ref{fig:phi_mu} and
\ref{fig:phi_mu2}. The initial increase in strength of the ratchet effect for
increasing values of $\lambda$ can be understood by considering the average
effective field
\begin{equation}
    \fat{B}^\mathrm{eff} = \fat{B}^\mathrm{ext} + \frac{1}{M} \sum_{i=1}^M \sum_{j \neq i} \fat{B}_{ij}^\mathrm{dip}
\end{equation} 
felt by the particles, where $M$ is the number of particles considered and
\begin{equation}
    \fat{B}_{ij}^\mathrm{dip}
    = 
    \frac{3 \fat{r}_{ij} (\fat{r}_{ij} \cdot \fatsym{\mu}_j)}{r^5}
    - \frac{\fatsym{\mu}_j}{r^3}.
\end{equation} 
This effective field is depicted in Fig.~\ref{fig:beff} for particles with
$\lambda = 1.44$, $T^* = 1$ and $\lambda = 2.45$, $T^* = 0.2$. These
temperatures and coupling strengths roughly correspond to the maxima in the net
torque and the averaged traversed angle [see Fig.~\ref{fig:phi_mu}]. The plots
in Fig.~\ref{fig:beff} show that, due to the dipolar interactions, the
effective field components $B^{\mathrm{eff}*}_x$ and $B^{\mathrm{eff}*}_y$ are
increased as compared to the components of the external field. The fact that an
enhancement of the (effective) field acting on the particles can support the
ratchet effect, is already suggested by our results for a non-interacting
system in Sec.~\ref{sec:ratchet}: As shown in Figs.~\ref{fig:torques_delta} and
\ref{fig:temp_torque}, an increase in $B_x$ alone or in both, $B_x$ and $B_y$,
can lead to larger values of the net torque.

\begin{figure}[ht]
    \centering
    \includegraphics[width=83mm]{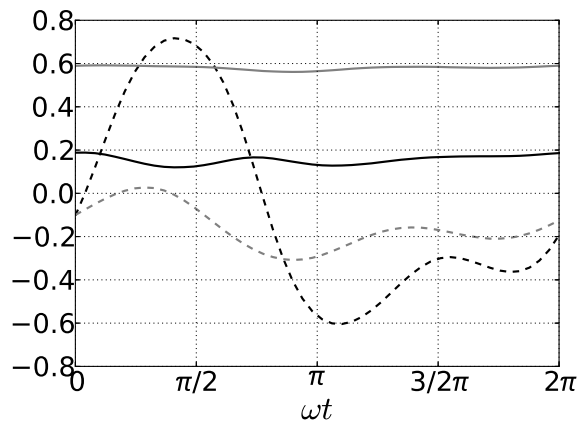}
    \caption{Mean orientation of the particles during one rotational period of
    the field for $\lambda = 1.44$ (black lines) and $\lambda = 9$ (gray
    lines). The $x$- and $y$-components of $\bar{\fat{S}}$ are indicated by the
    solid and dashed lines, respectively.} Here, $B_x^+ = 0.5$ and $T^* = 1$.
    \label{fig:exey13}
\end{figure}

For coupling strengths higher than $\lambda \approx 1.5$ ($T^* = 1$) or
$\lambda \approx 2.5$ ($T^* = 0.2$), respectively, the magnitude of the ratchet
effect (as measured by $\langle \bar{T}_z^* \rangle$) begins to decrease. We
relate this behavior to the increase of the ratio of conservative torques
induced by the dipolar interactions and the external field relative to the
strength of the noise. However, in contrast to the (corresponding) decrease
described in Sec.~\ref{sec:ratchet} [see Fig.~\ref{fig:temp_torque}], a large
contribution to the torque now stems from the dipole-dipole interaction and not
from the particle-field interaction. This can be seen in Fig.~\ref{fig:exey13},
where we compare the functions $\bar{S}_x(t)$ and $\bar{S}_y(t)$
[cf.~Eq.~(\ref{eq:mean_orient})] for a system at $\lambda = 1.44$ ($T^* = 1$)
and a more strongly coupled one at $\lambda = 9$ ($T^* = 1$).
It is seen that $\bar{S}_x$ is significantly larger for the latter system
($\bar{S}_x \approx 0.58$) than for the former one ($\bar{S}_x \approx 0.16$).
Recalling the discussion in Sec.~\ref{sec:ratchet} (non-interacting system),
one would thus expect the ratchet effect at $\lambda = 9$ to be even larger
than at $\lambda = 1.44$. However, the amplitude of $\bar{S}_y$ is considerably
smaller. Indeed, the maximum of $|\bar{S}_y|$ for the strongly coupled system
is about $0.31$, while it is about $0.72$ for the one with $\lambda = 1.44$. 

This means that the particles at $\lambda = 9$ are much more aligned along the
$x$-direction (i.e., the constant part of the field) without closely following
the oscillations in $y$-direction. In conclusion, the behavior seen at $\lambda
= 9$ is in stark contrast to what is shown in Fig.~\ref{fig:exey00} for a
non-interacting system. There, an increase in $B_x^*$ does not automatically
damp out the oscillations in the $y$-direction. Consequently, the ratchet
effect is increased rather than damped.

Finally, we note that, for relevant values of $\lambda$, the relative increase
in the net torque is larger for the low-temperature system than for the
high-temperature one. Indeed, Fig.~\ref{fig:phi_mu} shows that the value of
$T^*$ influences the entire behavior of $\langle \bar{T}_z^* \rangle$ as a
function of $\lambda$. Therefore, not only the dipolar, but also the
short-range repulsive interactions between the particles have an impact on the
ratchet effect.

\begin{figure}[ht]
    \centering
    \includegraphics[width=83mm]{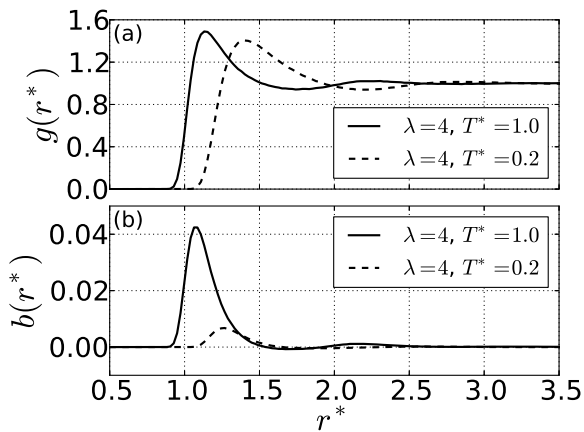}
    \caption{(a) Pair correlation functions of systems at different
    temperatures ($T^* = 1$, $0.2$) but identical dipolar coupling strengths
    $\lambda$. (b) The local field $b(r)$ [see Eq.~(\ref{eq:br})] for the two
    different systems.}
    \label{fig:bgr}
\end{figure}
The sensitivity against $T^*$ can be explained by the fact that the soft-sphere
interactions affect the effective distance between the dipolar particles. In
Fig~\ref{fig:bgr}(a), the radial distribution functions of two systems at
identical coupling strength $\lambda = 4$ but different temperatures ($T^* = 1$
and $0.2$) are shown. Judging from the position of the main peak, two neighbors
are typically closer to one another in the $T^* = 1$ system than in the low
temperature one. As a result the particles experience a considerably stronger
effective field in the high-temperature system. This is illustrated by
Fig.~\ref{fig:bgr}(b), where we plot the function
\begin{equation}
    \label{eq:br}
    b(r) = \frac{1}{S_r} \int_{S_r} \D S
    \frac{1}{N} \Bigg< \sum_{i=1}^N \sum_{j \neq i} \delta(\fat{r} - \fat{r}_{ij}) B_{x, ij}^\mathrm{dip} \Bigg>,
\end{equation}
where $S_r$ is the surface of a sphere of radius $r$. From a physical point of
view, the function $b(r)$ corresponds to the local field in $x$-direction that
is generated by neighboring dipolar particles with distance $r^*$ from the
central one. From Fig.~\ref{fig:bgr}(b), we see that the local field at short
distances is significantly increased in the $T^* = 1$ system as compared to the
field in the $T^* = 0.2$ system. Hence, as argued above, the rotational motion
is much more restricted in the former system resulting in a less pronounced
ratchet effect at a fixed coupling strength.

\subsection{Relation to self-assembly}
It is well established that strongly coupled dipolar particles can
self-assemble into a variety of structures including chains, networks, and
sheets \cite{jaeger_klapp, Jordanovic2011}. Moreover, for dense systems of
dipolar spheres, theory and computer simulations predict a phase with
spontaneous long-range, parallel (i.e., ferromagnetic) order \cite{klapp2002}.
It is therefore an interesting question whether these phenomena have any
relevance in the context of the rotational ratchet effect.

\begin{figure}[ht]
    \centering
    \includegraphics[width=72mm]{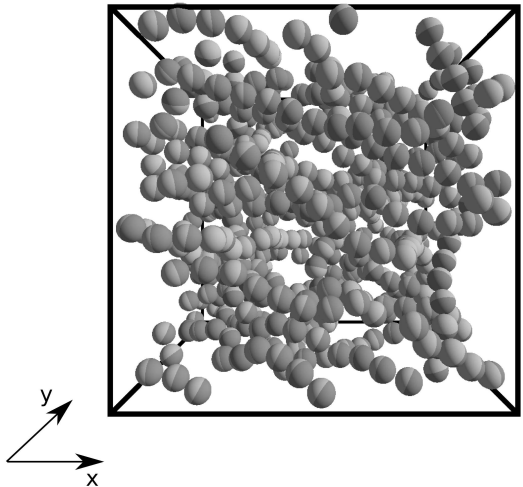}
    \caption{Snapshot of a system at $\lambda = 9$, $T^* = 1$, and $B_x^+ = 0.5$.}
    \label{fig:ss}
\end{figure}

The answer from our present BD simulations is essentially negative. Indeed, at
the conditions  where we found an increase in the ratchet effect ($T^* = 1$, $0
\lesssim \lambda \lesssim 3.5$; $T^* = 0.2$, $0 \lesssim \lambda \lesssim 5.5$)
there is no global parallel order. Moreover, significant local ordering of the
particles only occurs for dipolar coupling strengths $\lambda \gtrsim 9$ ($T^*
= 1$), which is outside of the range where we observe enhancement of the
ratchet effect. The structures seen in such a highly coupled systems are
illustrated by the simulation snapshot in Fig.~\ref{fig:ss}. Similar to
ferrofluids subject to constant, homogeneous external fields, the oscillating
field favors chain formation of the particles. Systems of lower coupling
strength lack any such order. In particular, no local order can be observed at
$\lambda = 1.44$, $T^* = 1$, i.e., where the ratchet effect is maximal.

Another interesting aspect is the (possible) impact of the ferromagnetic phase
transition occurring at higher densities. In Ref.~\cite{becker2007}, this
question was investigated on a mean-field level, where the particles experience
a (mean-field) torque of the form
\begin{equation}
    \label{eq:mf_torque}
    \fat{T}^\mathrm{mf}_i = \fat{e}_i \times \frac{K}{N} \sum_{i=1, j \neq i}^N \fat{e}_j,
\end{equation}
where $K$ is the coupling strength. For sufficiently large $K$, one finds a
spontaneous ferromagnetic ordering of the particles resulting in an effective
non-vanishing field component in $x$-direction \cite{becker2007}. Due to this
net field, the ratchet effect can occur even in the absence of an external
$x$-component of the field.

We have searched for a similar phenomenon in our many-particle system. However,
at the parameters considered, we were not able to find a net particle rotation.
Not even at high densities, where the dipolar soft spheres undergo a
ferromagnetic phase transition \cite{klapp2002}, did we detect such a rotation.

This could be due to several reasons: First, the true effective field within
the ferromagnetic phase is inhomogeneous and typically much weaker than any
average ``mean-field'' (this is also the reason that the
isotropic-ferromagnetic transition in a true dipolar system occurs at much
larger coupling strengths than those predicted by mean-field theory
\cite{klapp2002,becker2007}).

Second, in a dense dipolar system, the orientations of the particles are
strongly coupled over large distances. In other words, the dipole orientations
are severely restricted, which further suppresses the ratchet effect.

\section{Conclusions}
In this study, we have investigated the rotational thermal ratchet effect for
non-interacting particles as well as particles interacting via long-range
dipolar interactions.

With our particle based simulations, we looked at the angular trajectories of
the dipolar particles, which conclusively illustrate the net rotating behavior
of the driven particles. For non-interacting particles, we found that a finite
ratio of deterministic torques to random noise yields a maximally pronounced
ratchet effect.

The main focus of this study, however, was the investigation of the influence
of dipolar interactions on the rotational behavior of the particles. In
particular, we showed that dipolar interactions can have an enhancing as well
as a dampening effect depending on the dipolar coupling strength $\lambda$. The
enhancement found at small values of $\lambda$ is due to the fact that the
effective field acting on a particle is larger (than without interactions), but
not too large to suppress rotations. This finding is consistent with the
mechanism found in Ref.~\cite{becker2007}.

Interestingly, we were not able to attribute the increase in the ratchet effect
in systems of dipoles to a synchronization phenomenon, i.e., a coupled rotation
of two neighboring dipolar particles. It is, however, possible that such
synchronization phenomena occur at thermodynamic and field parameters that
differ from the ones investigated here.

At higher values of $\lambda$, i.e., stronger dipolar couplings, we find a
decrease in the ratchet effect. In this region, the particles start to
aggregate into clusters along the direction determined by the constant
contribution to the external field. As a consequence, the effective field
becomes too strong and the dipole moments can follow the oscillatory motion of
the field less and less, leading to a pronounced dampening of any rotations. We
note that the values of $\lambda$ considered in this work are in the range
typical for real ferrofluids (as are the considered densities).

As a somewhat counterintuitive effect, we have found that not only the
anisotropic dipolar interactions but also the isotropic repulsive interactions
between the particles have a significant influence on the ratchet effect. At
constant dipolar coupling strength, the steepness of these interactions
determines the average distance between the particles and thus, the magnitude
of the effective local field. In this way, short-range interactions can
``tune'' the effective torque.

In summary our results show that the conservative interactions typical of real
ferrofluids strongly influence noise-induced phenomena such as the ratchet
effect. So far, we have not taken into account the fact that the solvent, which
is omnipresent in a ferrofluid, induces additional hydrodynamic interactions
between the magnetic particles. These long-range interactions have been shown
to play a significant role in translational ratchets (see, e.g.,
Refs.~\cite{grimm2011,malgaretti2012}) and related synchronization phenomena
\cite{kotar2010}. The interplay of hydrodynamic and dipolar interactions in the
context of the present ratchet effect will be the subject of a future study.

\acknowledgments
We thank A. Engel for the motivation of this work and for helpful discussions.
Financial support from the DFG within the RTG 1558 {\em Nonequilibrium
Collective Dynamics in Condensed Matter and Biological Systems}, project B1, is
gratefully acknowledged.

\appendix*
\section{Reduced units}
Here, we show how the reduced units used in our study are related to the ones
used by Engel {\it et al.} \cite{engel2004}. The latter are denoted by a
superscript ``$\dagger$''. For the reduced temperature we find
\begin{equation}
    T^* = \mu^*_0 D^\dagger \frac{B^*}{B^\dagger}.
\end{equation}
The frequency is related by
\begin{equation}
    \omega^* = \frac{3}{D^\dagger},
\end{equation}
the time by
\begin{equation}
    t^* = \frac{1}{3} D^\dagger t^\dagger,
\end{equation}
and the torque by
\begin{equation}
    T^*_z = T^\dagger_z \frac{T^*}{D^\dagger}.
\end{equation}
Using $D^\dagger = 0.2$, $\mu^*_0 = 1$ [cf.~Eq.~(\ref{eq:Text})], and choosing
$B^\dagger = B^*$ and $\Delta t^\dagger = 0.0015$, yields $T^* = 0.2$, $\Delta
t^* = 0.0001$, and $\omega^* = 15$. This means that $T^*_z = T^\dagger_z$.


%


\end{document}